# Turnkey Technology: A Powerful Tool for Cyber Warfare

Mohamed Aly Bouke [1, *], Azizol Abdullah

[1]Department of Communication Technology and Network, Faculty of Computer Science and Information Technology, Universiti Putra Malaysia, Serdang 43400, Malaysia.

*Corresponding author(s). E-mail(s): bouke@ieee.org;
Contributing authors: azizol@upm.edu.my;

**Abstract**: Turnkey technology has emerged as a game-changing tool in cyber warfare, offering state and non-state actors unprecedented access to advanced cyber capabilities. The advantages of turnkey technology include rapid deployment and adaptability, lower costs and resource requirements, the democratization of cyber warfare capabilities, and enhanced offensive and defensive strategies. However, turnkey technology also introduces significant risks, such as the proliferation of cyber weapons, ethical considerations, potential collateral damage, escalation of conflicts, and legal ramifications. This paper provides a unique perspective on the implications of turnkey technology in cyber warfare, highlighting its advantages, risks, and challenges, as well as the potential strategies for mitigating these concerns. The research's novelty lies in examining real-world examples and proposing a multifaceted approach to address the challenges associated with turnkey technology in cyber warfare. This approach focuses on developing effective cybersecurity measures, establishing international norms and regulations, promoting responsible use and development of turnkey technology, and enhancing global cooperation on cyber warfare issues. By adopting this accountable and balanced approach, governments, organizations, and the international community can work together to create a more secure and stable digital environment, leveraging the benefits of turnkey technology while minimizing the associated risks and challenges.

1. **Introduction**

The digital revolution has given rise to a new era of warfare, where cyber-attacks and digital strategies play a crucial role in determining the outcomes of conflicts. Turnkey technology has emerged as a powerful tool in this landscape, providing state and non-state actors with advanced capabilities to conduct cyber warfare [1]. Turnkey technology refers to pre-built, ready-to-use software and hardware systems that can be easily implemented with minimal customization or setup. These systems are designed to provide users with a comprehensive solution that can be quickly deployed to achieve specific objectives. Turnkey technology encompasses many tools and systems in cyber warfare, including malware, exploit kits, command and control servers, and even full-scale offensive cyber capabilities [2]. These technologies can be acquired off-the-shelf, allowing users to quickly and efficiently execute cyber attacks or defensive measures without the need for extensive technical expertise or resources.

Cyber warfare has evolved significantly since its inception. In the early days, cyber attacks were often limited to basic hacking attempts and distributed denial-of-service (DDoS) attacks. However, as technology advanced and the internet became an integral part of everyday life, cyber warfare began to take on new forms. State-sponsored cyber attacks and espionage emerged, targeting critical infrastructure, political organizations, and military systems. High-profile attacks, such as Stuxnet, which targeted Iranian nuclear facilities, and the NotPetya ransomware, which caused widespread damage globally, have demonstrated the growing sophistication and scale of cyber warfare[3,4].

The rise of turnkey technology has played a significant role in this evolution, enabling a more comprehensive range of actors to engage in cyber warfare. By providing an easily accessible means to launch cyber attacks, turnkey technology has effectively democratized cyber warfare capabilities, creating new opportunities and challenges for state and non-state actors [5,6].

---

* Corresponding Author: bouke@ieee.org

The increasing prevalence of turnkey technology in cyber warfare has had significant implications for modern conflicts. The rapid deployment and adaptability of turnkey systems make them an attractive option for states and non-state actors seeking to project power and influence in the digital realm. As a result, turnkey technology has become an integral component of many military and intelligence operations, as well as asymmetric warfare strategies employed by smaller nations and non-state actors [7,8].

Furthermore, the low cost and ease of access to turnkey technology have led to the proliferation of cyber warfare capabilities across the globe. This has heightened the potential for cyber conflict between nations and created new risks and vulnerabilities, as non-state actors such as terrorist organizations, hacktivists, and criminal networks can now exploit these technologies for their purposes. Consequently, integrating turnkey technology into modern conflicts has introduced new dimensions to warfare and necessitates a comprehensive understanding of its implications for global security and stability.

The rest of this paper is organized as follows: Section 2 presents a comprehensive literature review, providing an overview of turnkey technology and its role in cyber warfare. Section 3 delves into real-world examples of turnkey technology as a cyber warfare weapon, showcasing the diverse nature of its application. Section 4 offers a critical discussion and analysis of the examples, highlighting the key factors that differentiate them and the implications of using turnkey technology in cyber warfare. Finally, Section 5 concludes the paper by summarizing the main findings and emphasizing the importance of understanding and preparing for the evolving landscape of cyber warfare as turnkey technology continues to advance and become more accessible.

## 2. Literature Review

The emergence of turnkey technology in cyber warfare has garnered significant attention in academic and policy-making circles. As the threat posed by turnkey technology in cyber warfare grows, researchers emphasize the importance of developing effective countermeasures to protect critical infrastructure and prevent catastrophic cyber-attacks. This literature review explores key works in the field, focusing on turnkey technology development, its application in cyber warfare, and potential countermeasures to mitigate associated risks.

Cavelty [9] explores the concept of cyber-power within the European Union's (EU) cyber-security policy. It highlights the multifaceted nature of cyber-power and proposes two notions: (1) a narrow understanding focusing on a political entity's "cyber-fitness," and (2) a broader understanding that views cyber-power as an auxiliary type of power for internal and external political purposes. The authors emphasize the need for future research to examine both aspects and their implications for the EU's role in global politics.

Jensen [10] examines the challenges that NATO faces in coordinating offensive cyberspace operations capabilities (OCOC) among its members. Due to the sensitive and secretive nature of OCOCs, NATO members have legitimate interests in maintaining secrecy around these capabilities, resulting in a dilemma for the alliance. The article argues that while NATO's current limited approach to OCOC coordination is pragmatic, real-world inhibitions to information sharing may be moderated by factors such as external threats and the evolving maturity of the cyber domain.

Turnkey technology, also known as "off-the-shelf" or "plug-and-play" technology, refers to pre-packaged solutions that can be easily deployed with minimal customization or technical expertise [11]. The adoption of turnkey technology in cyberwarfare has grown in recent years, as evidenced by the increasing number of cyber-attacks using these tools [12]. Kostyuk and Zhukov [12] argue that the proliferation of turnkey technology has lowered the barriers to entry for state and non-state actors, enabling even those with limited resources to conduct sophisticated cyber-attacks.

The democratization of cyberwarfare capabilities is a significant concern among security experts [13]. With the availability of turnkey technology, small states, and non-state actors can potentially access advanced cyber capabilities previously reserved for powerful nation-states [14]. Rid, and Buchanan [15] argue that this development may lead to a more volatile cyber landscape as actors with different motivations and objectives gain access to powerful cyber weapons.

The rise of turnkey technology in cyberwarfare has significant implications for the strategies employed by state and non-state actors. Smeets [16] posits that the proliferation of turnkey technology may also lead to increased deterrence as states become more aware of their vulnerabilities and improve their cyber defenses.

In conclusion, this literature underscores the increasing risks associated with democratizing cyber capabilities and their potential consequences for national security and international stability. Although researchers have suggested several countermeasures to alleviate these risks, additional research is required to fully comprehend the evolving cyber landscape and identify the most effective strategies to tackle the threats posed by turnkey technology.

3. **Turnkey Technology as Cyber Warfare Weapon – Real-World Examples**

This section will explore real-world examples of turnkey technology used as a cyber warfare weapon. These cases were selected based on their relevance to understanding the role of turnkey technology in cyber warfare, its potential risks, and its impact on the balance of power in cyberspace. Additionally, these examples demonstrate the increasing accessibility and effectiveness of turnkey solutions, which enable even less sophisticated actors to launch potent cyber attacks. We will critically analyze each example aspect concerning our paper's objectives, which include understanding the role of turnkey technology in cyber warfare, its potential risks, and its impact on the balance of power in cyberspace.

**3.1. Example 1: Stuxnet**

Stuxnet, a highly sophisticated computer worm discovered in 2010, targeted Iran's nuclear facilities and disrupted their uranium enrichment program [17]. Widely believed to be a joint operation between the United States and Israel, Stuxnet is a prime example of turnkey technology in cyber warfare [18–20].

*Turnkey Aspects of Stuxnet:*

- Readily Deployable: Stuxnet demonstrates how turnkey technology can be employed as a ready-to-use cyber warfare tool. Its complexity and sophistication allowed it to be deployed quickly and with minimal modifications, effectively evading detection and causing significant damage to Iran's nuclear program. This ease of deployment exemplifies the potential for turnkey technology to be utilized for swift and targeted cyber attacks.
- Specialized Knowledge: The developers of Stuxnet possessed specialized knowledge of the target systems, including the industrial control systems (ICS) used in Iran's nuclear facilities. This expertise enabled them to design a turnkey solution tailored specifically to the target environment, increasing the effectiveness of the attack.
- Exploiting Zero-Day Vulnerabilities: Stuxnet's use of four zero-day vulnerabilities in Microsoft Windows illustrates the potential for turnkey technology to exploit unknown vulnerabilities, making it challenging for defenders to protect against such attacks. This aspect of turnkey technology enhances its potency as a cyber warfare tool.
- Adaptable and Proliferable: Other threat actors could be reverse-engineered and adapt Stuxnet's techniques and components, making advanced cyber warfare capabilities more accessible. This adaptability highlights the potential for turnkey technologies to increase and be employed by a broader range of actors, increasing the risks associated with cyber warfare.
- Strategic Impact: Stuxnet had significant geopolitical implications, demonstrating the potential of turnkey technologies to be employed as powerful weapons in achieving strategic objectives. Its success in causing considerable damage to Iran's nuclear program underscores the effectiveness of turnkey technology in cyber warfare and highlights its potential to influence global power dynamics.
- Countermeasures and Lessons Learned: Stuxnet's discovery led to greater scrutiny of industrial control systems security and an increased focus on securing critical infrastructure. This case underscores the importance of continuously improving cyber defenses and sharing information about vulnerabilities and threats among organizations and governments.
- Ethical Considerations: Stuxnet raises questions about the ethics of using cyber weapons to target critical infrastructure, as the potential for unintended consequences and collateral damage is high. It also highlights the importance of establishing norms and agreements for state behavior in cyberspace to prevent conflict escalation and ensure the digital domain's stability.

Through a critical analysis of Stuxnet's turnkey aspects, it becomes evident that this cyber weapon epitomizes the potential of turnkey technology in cyber warfare. The ready-to-use nature of Stuxnet, coupled with its specialized knowledge, exploitation of zero-day vulnerabilities, adaptability, and strategic impact, underscores the power and risks associated with turnkey technology in cyber warfare.

### 3.2. Example 2: WannaCry Ransomware

WannaCry, a massive ransomware attack in May 2017, impacted thousands of organizations and individuals across the globe [21]. The attack targeted numerous sectors, including healthcare, finance, and government institutions. By exploiting a vulnerability in the Windows operating system, WannaCry encrypted the victims, rendering them inaccessible. The attackers then demanded ransom payments in Bitcoin to release the decryption keys, causing significant financial losses and operational disruptions [22,23].

*Turnkey Aspects of WannaCry:*

- Readily Deployable: WannaCry demonstrates the potential for turnkey technology to be employed as a ready-to-use cyber warfare tool. The ransomware was designed to spread through networks quickly and encrypt data on infected systems, requiring minimal interaction from the attacker. This ease of deployment exemplifies the potential for turnkey technology to enable swift and widespread cyber attacks.
- Exploiting Stolen Cyber Weapons: WannaCry incorporated a powerful cyber weapon known as EternalBlue, allegedly stolen from the US National Security Agency (NSA) and leaked by a group called the Shadow Brokers. This aspect highlights the potential risks associated with turnkey technology. It demonstrates how cyber weapons developed by nation-states can be repurposed and utilized by other actors, making advanced cyber warfare capabilities more accessible.
- Ransomware as a Service (RaaS): WannaCry is an example of Ransomware as a Service, a turnkey technology that enables even relatively inexperienced cybercriminals to launch ransomware attacks. The availability of RaaS platforms has lowered the barriers to entry for cyber warfare, contributing to the proliferation of ransomware attacks and increasing the risks associated with cyber warfare.
- Global Impact: WannaCry had a widespread impact, affecting organizations and individuals in over 150 countries, including hospitals, government institutions, and businesses. This demonstrates turnkey technology's potential to cause significant global damage and its capacity to influence the balance of power in cyberspace.
- Attribution Challenges: The WannaCry attack raised significant challenges in attributing responsibility, as the ransomware was developed using a stolen cyber weapon, and its origin was initially unclear. This highlights the potential for turnkey technology to be employed by state and non-state actors to conduct cyber warfare operations while obfuscating their involvement, complicating the dynamics of cyber warfare.
- Countermeasures and Lessons Learned: The WannaCry attack prompted organizations worldwide to reevaluate their cybersecurity practices, particularly regarding patch management and system updates. This case highlights the importance of implementing proactive security measures and maintaining up-to-date systems to mitigate the risks posed by turnkey technology.
- Ethical Considerations: WannaCry raises ethical concerns regarding nation-states' development and use of cyber weapons and the potential for these tools to be repurposed and used by other malicious actors. The attack highlights the need for governments to be more responsible in managing their cyber arsenals and sharing vulnerability information with the broader cybersecurity community.

The examination of this example reveals that this ransomware attack exemplifies the potential of turnkey technology in cyber warfare. The ready-to-use nature of WannaCry, combined with its exploitation of stolen cyber weapons, the RaaS model, global impact, and attribution challenges, underscores the power and risks associated with turnkey technology in cyber warfare.

### 3.3. Example 3: NotPetya Attack

The NotPetya attack in June 2017 was a highly destructive cyber attack that targeted businesses, government institutions, and critical infrastructure worldwide [4]. Initially believed to be a ransomware attack similar to WannaCry, it was later determined to be a wiper malware with the primary objective of causing widespread damage and disruption. NotPetya is attributed to the Russian government and targeted Ukraine most heavily, affecting the country's government, banks, and power grid [24–26].

**Turnkey Aspects of NotPetya:**

- Rapid Propagation: NotPetya demonstrated the ability of turnkey technology to spread across networks and rapidly inflict damage on targeted systems. The malware used worm-like propagation methods and credential-stealing techniques, enabling it to infect many systems quickly and efficiently. This rapid propagation is a crucial characteristic of turnkey technology in cyber warfare, allowing for swift and widespread attacks.
- Weaponizing Legitimate Tools: To further its objectives, NotPetya weaponized legitimate software tools, such as the system management tool PsExec and the credential-dumping tool Mimikatz. This highlights the risks of turnkey technology in cyber warfare, as it demonstrates how existing tools can be repurposed and combined with malicious intent to create devastating cyber weapons.
- Destructive Impact: Unlike typical ransomware attacks, NotPetya was designed primarily to cause damage and disruption. The malware encrypted the victim's data and made it nearly impossible to recover, resulting in significant financial losses and operational disruptions. This showcases the potential of turnkey technology to be utilized for destructive purposes in cyber warfare, causing a widespread and lasting impact on targeted systems and organizations.
- Difficult Attribution: The attribution of the NotPetya attack to the Russian government was challenging due to its use of multiple obfuscation techniques, such as using legitimate tools and the ransomware disguise. This highlights another risk of turnkey technology in cyber warfare, as it enables state and non-state actors to conduct cyber operations while hiding their involvement, complicating the dynamics of cyber warfare.
- Geopolitical Implications: The NotPetya attack had significant geopolitical implications, as it was attributed to the Russian government and primarily targeted Ukraine, exacerbating existing tensions between the two countries. This demonstrates the potential of turnkey technology to be used as a strategic tool in cyber warfare, influencing the balance of power in cyberspace and contributing to geopolitical instability.
- Countermeasures and Lessons Learned: The NotPetya attack underscored the importance of robust incident response planning and the need for organizations to invest in cybersecurity resilience. This case serves as a reminder that organizations must be prepared for potential cyber-attacks and have the necessary resources and plans to recover quickly and efficiently.
- Ethical Considerations: NotPetya raises ethical questions regarding nation-states' use of destructive cyber weapons, especially when they can potentially cause significant collateral damage to non-targeted entities. This case emphasizes the need for clear rules of engagement and international agreements to prevent the escalation of cyber conflicts and to limit the harm caused by cyber warfare.

The rapid propagation, weaponization of legitimate tools, destructive impact, difficult attribution, and geopolitical implications of NotPetya highlight the power and risks associated with turnkey technology in cyber warfare. By critically analyzing the turnkey aspects of the NotPetya attack, we can see that this event provides valuable insights into turnkey technology's role, risks, and impact in cyber warfare.

### 3.4. Example 4: DDoS-for-Hire Services

Distributed Denial of Service (DDoS) attacks have been a prevalent form of cyber attack for decades. In recent years, however, DDoS-for-hire services have emerged as a turnkey technology in cyber warfare, making it easier for individuals and organizations to launch DDoS attacks without technical expertise or resources [27,28]. These services can be bought on the dark web or through more accessible online marketplaces, allowing virtually anyone to launch a DDoS attack with minimal effort [29,30].

*Turnkey Aspects of DDoS-for-Hire Services:*

- Accessibility and Ease of Use: DDoS-for-hire services have democratized access to powerful cyber warfare tools, enabling individuals and organizations with limited technical skills to conduct disruptive DDoS attacks. This highlights the role of turnkey technology in cyber warfare by making advanced cyber capabilities accessible to a broader range of actors.
- Low Cost: DDoS-for-hire services are often inexpensive, making them an attractive option for individuals or organizations with limited resources who wish to conduct cyber attacks. This low cost further illustrates the role of turnkey technology in cyber warfare by lowering the barriers to entry for potential attackers.
- Amplification of Attacks: DDoS-for-hire services often use amplification techniques to increase traffic directed at a target, making the attack more disruptive and difficult to mitigate. This amplification highlights the risks associated with turnkey technology in cyber warfare, as it can enable small-scale actors to inflict significant damage on their targets.
- Anonymity: DDoS-for-hire services can provide a degree of anonymity to the attacker, as the service acts as an intermediary, making attribution more challenging. This anonymity further emphasizes the risks associated with turnkey technology in cyber warfare, as it enables attackers to conduct operations with reduced risk of retaliation or legal consequences.
- Destabilizing Effect: The widespread availability and use of DDoS-for-hire services can destabilize the balance of power in cyberspace. With more actors able to launch disruptive DDoS attacks, the likelihood of conflict and escalation in cyberspace increases, contributing to a more volatile and unpredictable environment.
- Countermeasures and Lessons Learned: The rise of DDoS-for-hire services highlights the need for organizations to improve their DDoS mitigation strategies and invest in technologies that can effectively handle these types of attacks. This case also underscores the importance of international cooperation and law enforcement efforts to crack down on the providers of these services.
- Ethical Considerations: DDoS-for-hire services raise ethical concerns about the ease with which individuals and organizations can conduct disruptive cyber attacks, often with little regard for the consequences or the harm caused to innocent parties. This case emphasizes the need for greater accountability in cyberspace and the development of legal frameworks to hold malicious actors responsible for their actions.

The accessibility, low cost, amplification, anonymity, and destabilizing effect of DDoS-for-hire services underscore turnkey technology's potential dangers and complexities in cyber warfare. By critically analyzing the turnkey aspects of DDoS-for-hire services, we can see that this phenomenon provides valuable insights into turnkey technology's role, risks, and impact in cyber warfare.

### 3.5. Example 5: China's 5G (Huawei)

China's 5G technology, particularly the advancements made by Huawei, has become a point of contention between China and Western countries such as the United States and the European Union [31]. Western countries have expressed concerns about the security risks associated with using Huawei's 5G infrastructure, fearing that it could be used for cyber espionage and cyber warfare [32].

**Turnkey Aspects of China's 5G (Huawei):**

- Strategic Advantage: China's 5G technology, if integrated into global communication networks, could potentially give China a strategic advantage in cyber warfare, as it would have access to critical infrastructure and sensitive data. This aspect highlights turnkey technology's potential risks and impact in cyber warfare.
- Attribution Complexity: The use of 5G technology in cyber warfare could potentially increase attribution complexity, as the source of an attack may be difficult to trace, particularly if it originates from a country with advanced cyber capabilities. This further emphasizes the challenges posed by turnkey technology in cyber warfare.
- Economic and Political Leverage: The global race to deploy 5G technology has significant economic and political implications. The concerns over Huawei's 5G technology have led to a push for alternative infrastructure providers, which could impact the balance of power in cyberspace and shape future cyber warfare strategies.
- Countermeasures and Lessons Learned: The concerns surrounding China's 5G technology highlight the importance of understanding the potential risks associated with turnkey technology in cyber warfare. Governments and organizations must develop comprehensive strategies to secure their communication networks and mitigate potential threats.
- Ethical Considerations: The debate surrounding China's 5G technology raises ethical questions about the role of private companies in cyber warfare and the potential risks associated with integrating foreign technology into national infrastructure. This case underscores the need for a broader discussion on the ethical implications of using turnkey technology in cyber warfare and the development of international norms and regulations to govern its use.

### 3.6. Analysis and comparisons of the real-world examples

In this section, we critically analyzed the use of turnkey technology as a cyber warfare tool, focusing on five real-world examples: Stuxnet, WannaCry Ransomware, NotPetya Attack, DDoS-for-Hire Services, and China's 5G (Huawei). To better understand and compare these examples, we included a table that presents critical factors such as Attribution Complexity, Impact on Infrastructure, Level of Expertise Required, and Potential for Escalation.

Table 1 Comparison of Key Turnkey Technology Cyber Warfare Examples.

| Example | Attribution Complexity | Impact on Infrastructure | Level of Expertise Required | Potential for Escalation |
| --- | --- | --- | --- | --- |
| Stuxnet | High | High | High | Moderate |
| WannaCry Ransomware | Moderate | High | Low | Low |
| NotPetya Attack | High | High | Moderate | Moderate |
| DDoS-for-Hire Services | Low | Moderate | Low | Low |
| China's 5G (Huawei) | Moderate | High | High | Moderate |

As seen in Table 1 above, the examples differ in various aspects, underscoring the versatility of turnkey technology in cyber warfare. Stuxnet demonstrates the potential for high attribution complexity, significant infrastructure impact, and the necessity for advanced expertise. However, it also highlights the risk of unintentional escalation, as the malware inadvertently spreads beyond its intended targets.

WannaCry Ransomware, on the other hand, showcases how less-skilled attackers can employ turnkey technology to cause substantial damage to infrastructure. Despite its moderate attribution complexity, the ransomware attack had a relatively low potential for escalation, as it was primarily financially motivated.

The NotPetya Attack shared similarities with Stuxnet and WannaCry regarding attribution complexity and infrastructure impact. However, it required a moderate level of expertise and had a reasonable potential for escalation, as the attack was primarily aimed at causing disruption rather than financial gain.

DDoS-for-Hire Services exemplify the accessibility of turnkey technology in cyber warfare. These services require minimal expertise, have low attribution complexity, and generally have moderate infrastructure impact. While they also have a low potential for escalation, their widespread availability increases the likelihood of such attacks occurring.

The concerns surrounding China's 5G technology, particularly the involvement of Huawei, illustrate the potential risks associated with turnkey technology in cyber warfare. Despite the company's denial of wrongdoing, the potential impact on infrastructure is high, given the pervasive nature of 5G networks. As the infrastructure necessary for providing advanced communication capabilities can also be perceived as a potential tool for cyber espionage or attacks, the attribution complexity, in this case is moderate.

The expertise required to exploit 5G technology for cyber warfare would be high, as it involves tampering with sophisticated communication equipment and bypassing security measures. The potential for escalation is also moderate. Countries concerned about the security risks associated with Huawei's 5G technology may resort to retaliatory measures or implement policies restricting its usage, increasing tensions between nations.

Addressing these concerns requires a collaborative effort among governments, private companies, and international organizations to establish trust, transparency, and common standards for Cybersecurity in next-generation communication technologies. This example further underscores the challenges associated with turnkey technology in cyber warfare and emphasizes the importance of international cooperation in addressing the related challenges.

In conclusion, comparing these real-world examples emphasizes the diverse nature of turnkey technology in cyber warfare. The cases demonstrate that various actors can utilize these tools, from highly skilled nation-states to relatively unskilled individuals. Moreover, they highlight the varying degrees of attribution complexity, infrastructure impact, and potential for escalation resulting from turnkey technology as a cyber warfare tool. As such, governments, organizations, and individuals must understand and prepare for the ever-evolving landscape of cyber warfare, particularly as turnkey technology advances and becomes more accessible. The need for international cooperation, establishing trust, and developing legal and ethical frameworks is crucial in addressing the challenges posed by the proliferation of turnkey technology in cyber warfare and ultimately contributing to a more secure and stable cyberspace environment.

## 4. Discussion

The democratization of cyber warfare tools through turnkey technology has resulted in a more volatile and unpredictable landscape. In this paper, we have delved into the role of turnkey technology in cyber warfare by analyzing various real-world examples and highlighting its potential risks, impact, and implications for the balance of power in cyberspace. This section provides a more detailed and critical discussion of the findings.

One of the most significant challenges posed by turnkey technology in cyber warfare is the rapid evolution and adaptation of these tools. With the rapid advancement of technology, threat actors can easily modify and update their turnkey tools to bypass security measures, making it difficult for defenders to keep up. This dynamic nature of turnkey technology, coupled with its ease of use, exacerbates the threats associated with cyber warfare.

Secondly, turnkey technology in cyber warfare has blurred the lines between state-sponsored and non-state actors, complicating attribution and response efforts. This ambiguity further complicates the development of effective cyber policies and international norms. In some cases, turnkey technology can be used as a smokescreen to hide the true intentions or identity of the attacker, making it difficult for governments and organizations to determine appropriate countermeasures and retaliatory actions.

Additionally, the widespread use of turnkey technology has created a situation where the level of expertise required to engage in cyber warfare has significantly decreased. This has led to a new breed of cybercriminals, referred to as "script kiddies," who have little to no technical knowledge but can still carry out sophisticated attacks using turnkey tools. This phenomenon has expanded the pool of potential threat actors, making it more challenging for organizations to protect against and respond to cyber threats.

Moreover, the proliferation of turnkey technology has raised concerns over the potential for cyber warfare to escalate into kinetic warfare. As these tools enable more actors to engage in cyber warfare, the likelihood of misunderstandings, misattributions, and unintended consequences increases. This, in turn, could potentially result in a retaliatory cycle that escalates beyond cyberspace and into the physical domain, increasing the risk of conflict between nations.

The growing reliance on turnkey technology in cyber warfare has also raised questions regarding the ethics of using such tools. The ease of access to these technologies may lead to a lower threshold for engaging in cyber warfare, potentially resulting in more frequent and indiscriminate attacks on civilian infrastructure and systems. This highlights the need for a broader discussion on the ethical implications of using turnkey technology in cyber warfare and the development of ethical guidelines and frameworks to guide the use of these tools.

Furthermore, the role of the private sector in developing and distributing turnkey technology is another critical aspect to consider. In many cases, private companies produce and distribute turnkey tools that can be repurposed for cyber warfare. This raises questions about the responsibility and accountability of these companies and the need for increased regulation and oversight to prevent the misuse of their products.

The importance of international cooperation in addressing the challenges posed by turnkey technology in cyber warfare cannot be overstated. Countries must establish common standards, share intelligence, and collaborate on developing effective defensive and offensive strategies to counter the risks associated with turnkey technology.

Finally, the potential for cyber warfare to become a more significant aspect of geopolitical conflict as turnkey technology becomes more accessible should not be overlooked. As turnkey technology proliferates, cyber warfare may become an increasingly prominent tool for nation-states to exert power and influence, leading to a greater need for international norms and agreements to prevent escalatory and destabilizing actions in cyberspace.

In conclusion, the extended discussion highlights the complexities and challenges associated with the widespread use of turnkey technology in cyber warfare. The democratization of cyber warfare tools, the blurring of lines between state-sponsored and non-state actors, the emergence of "script kiddies," the potential escalation of cyber warfare into kinetic warfare, the ethical implications of using turnkey technology, the role of the private sector in the development and distribution of turnkey tools, the importance of international cooperation, and the potential for cyber warfare to become a more significant aspect of geopolitical conflict are all critical aspects that need to be considered when addressing the challenges posed by the proliferation of turnkey technology in cyber warfare.

To tackle these challenges, governments, organizations, and individuals must develop a multifaceted approach, including policy development, technological innovation, international cooperation, and increased awareness of the risks and consequences of turnkey technology in cyber warfare. By doing so, we can better understand and prepare for the ever-evolving landscape of cyber warfare, particularly as turnkey technology advances and becomes more accessible.

5. **Conclusion**

In conclusion, this paper has provided a critical analysis of the role of turnkey technology in cyber warfare, emphasizing its potential risks, impact, and implications for the balance of power in cyberspace. We have demonstrated the widespread availability and use of turnkey technology in cyber warfare through real-world examples and have underscored the dangers and complexities associated with its proliferation.

As the cyberspace landscape continues to evolve and become more interconnected, governments, organizations, and cybersecurity professionals must stay vigilant and understand the role and implications of turnkey technology in cyber warfare. This understanding can help inform the development of effective strategies, policies, and technological innovations to address the challenges posed by turnkey technology, ultimately contributing to a more secure and stable cyberspace environment.

Future research could delve deeper into developing specific countermeasures and policy initiatives tailored to mitigate the risks associated with different types of turnkey technology in cyber warfare. Additionally, investigating the role of international cooperation, public-private partnerships, and establishing global cybersecurity norms could provide valuable insights into addressing the challenges posed by the proliferation of turnkey technology in cyber warfare. Exploring the ethical implications of turnkey technology and creating ethical guidelines for its use can also contribute to a more reliable and secure cyberspace. As turnkey technology advances, fostering a collaborative approach will be crucial in navigating the ever-changing cyber warfare landscape.

**Declarations**

**Funding statement**

Not applicable.

**Declaration of Interests Statement**

Conflict of Interests There is no conflict of interest between the authors.

**Data availability statement**

Data included in article/supplementary material/referenced in the article.

**Additional information**

No additional information is available for this paper.